\documentclass[a4]{emulateapj}

\shorttitle{MAGIC observations of the unidentified TeV $\gamma$-ray source TeV J2032+4130}
\shortauthors{Albert et al.}

\begin{document}

\title{MAGIC observations of the unidentified $\gamma$-ray source TeV J2032+4130}
\author{
 J.~Albert\altaffilmark{a}, 
 E.~Aliu\altaffilmark{b}, 
 H.~Anderhub\altaffilmark{c}, 
 P.~Antoranz\altaffilmark{d}, 
 C.~Baixeras\altaffilmark{e}, 
 J.~A.~Barrio\altaffilmark{d},
 H.~Bartko\altaffilmark{f}, 
 D.~Bastieri\altaffilmark{g}, 
 J.~K.~Becker\altaffilmark{h},   
 W.~Bednarek\altaffilmark{i}
% ...
}
 \altaffiltext{A} {Universit\"at W\"urzburg, D-97074 W\"urzburg, Germany}
 \altaffiltext{B} {IFAE, Edifici Cn., E-08193 Bellaterra (Barcelona), Spain}
 \altaffiltext{C} {ETH Zurich, CH-8093 Switzerland}
 \altaffiltext{D} {Universidad Complutense, E-28040 Madrid, Spain}
 \altaffiltext{E} {Universitat Aut\`onoma de Barcelona, E-08193 Bellaterra, Spain}
 \altaffiltext{F} {Max-Planck-Institut f\"ur Physik, D-80805 M\"unchen, Germany}
 \altaffiltext{G} {Universit\`a di Padova and INFN, I-35131 Padova, Italy}  
 \altaffiltext{H} {Universit\"at Dortmund, D-44227 Dortmund, Germany}
 \altaffiltext{I} {University of \L\'od\'z, PL-90236 Lodz, Poland} 
 \altaffiltext{J} {Universitat de Barcelona, E-08028 Barcelona, Spain}
 \altaffiltext{K} {Yerevan Physics Institute, AM-375036 Yerevan, Armenia}
 \altaffiltext{L} {Tuorla Observatory, Turku University, FI-21500 Piikki\"o, Finland}
 \altaffiltext{M} {Inst. de Astrofisica de Canarias, E-38200, La Laguna, Tenerife, Spain}
 \altaffiltext{N} {Universit\`a di Udine, and INFN Trieste, I-33100 Udine, Italy} 
 \altaffiltext{O} {Universit\`a  di Siena, and INFN Pisa, I-53100 Siena, Italy}
 \altaffiltext{P} {University of California, Davis, CA-95616-8677, USA}
 \altaffiltext{Q} {Humboldt-Universit\"at zu Berlin, D-12489 Berlin, Germany} 
 \altaffiltext{R} {Inst. for Nucl. Research and Nucl. Energy, BG-1784 Sofia, Bulgaria}
 \altaffiltext{S} {INAF/Osservatorio Astronomico and INFN, I-34131 Trieste, Italy} 
 \altaffiltext{T} {Universit\`a  di Pisa, and INFN Pisa, I-56126 Pisa, Italy}
 \altaffiltext{U} {ICREA \& Institut de Ci\`encies de l'Espai (IEEC-CSIC), 08193 Barcelona, Spain} 
 \altaffiltext{V} {Depto. de Astrofisica, Universidad, E-38206 La Laguna, Tenerife, Spain} 
 \altaffiltext{X} {deceased}
 \altaffiltext{Y} {Present address: APC (CNRS) Paris, France} 		
 \altaffiltext{\dag} {Corresponding authors}

\begin{abstract}
We observed the first known very high energy (VHE) $\gamma$-ray
emitting unidentified source, TeV J2032+4130, for 94 hours with the
MAGIC telescope. The source was detected with a significance of
5.6$\sigma$. The flux, position, and angular extension are compatible
with the previous ones measured by the HEGRA telescope system five
years ago. The integral flux amounts to (4.5$\pm$0.3$_{\rm
stat}\pm$0.35$_{\rm sys}$)$\times$10$^{-13}$ ph~cm$^{-2}$~s$^{-1}$
above 1~TeV. The source energy spectrum, obtained with the lowest
energy threshold to date, is compatible with a single power law with a
hard photon index of $\Gamma$=-2.0$\pm$0.3$_{\rm stat}\pm$0.2$_{\rm
sys}$.	
\end{abstract}


\section{Introduction}

The TeV source J2032+4130 (Aharonian et al. 2002) was the first
unidentified very high energy (VHE) $\gamma$-ray source, and also
the first discovered extended TeV source, likely to be Galactic.

Intensive observational campaigns at different wavelengths have been
carried out on TeV J2032+4130. Butt et al. (2003) presented an
analysis of the CO, HI, and infrared emissions, together with first
observations by {\it Chandra} (5 ksec) and a reanalysis of VLA
data. These observations showed that the TeV source region is
positionally coincident with an outlying group of stars (from the
Cygnus OB2 core), although they failed to identify a
counterpart. Mukherjee et al. (2003) analyzed the same {\it Chandra}
data and provided optical follow-up observations of several of the
brightest X-ray sources, confirming that most were either O stars or
foreground late-type stars. A deeper {\it Chandra} observation (50
ksec, Butt et al. 2006), found hundreds of star-like sources and yet
no diffuse X-ray counterpart emission.

A deep ($\sim$ 50~ksec) \textit{XMM-Newton} exposure has also been
obtained (Horns et al. 2007). After the subtraction of the
contribution of known sources from the data, an extended X-ray
emission region with a FWHM size of $\sim$~12 arcmin was reported. The
centroid of the emission is co-located with the position of TeV
J2032+4130 and was proposed as the counterpart of the TeV source. The
question whether the result reported by Horns et al. can be interpreted
as a truly diffuse background, or it could be a result of unresolved
X-ray sources, remains disputable.

Paredes et al. (2007) and Mart\'i et al. (2007) have provided deep
radio observations covering the TeV J2032+4130 vicinity using the
Giant Metrewave Radio Telescope and discovered a population of radio
sources, some in coincidence with X-ray detections by Butt et
al. (2006) and with optical/IR counterparts. At least three of these
sources are non-thermal, and one has a hard X-ray energy spectrum. They found
extended non-thermal diffuse emission in the radio band apparently
connecting with one or two radio sources. It is yet to be determined
if one or more of these sources is similar to some of the known
$\gamma$-ray binaries (e.g., Aharonian et al. 2006, Albert et
al. 2006a).

Several theoretical explanations for the TeV emission from J2032+4130
have been given. Among them, those related with extragalactic
counterparts, e.g., a radiogalaxy (Butt et al. 2006) or a proton
blazar (Mukherjee et al. 2003), face the difficulty of explaining the
extended appearance of the source. Gamma-ray production in
hypothetical jet termination lobes of Cyg X-3 was explored (Aharonian
et al. 2002), but the putative northern lobe of Cyg X-3 (now
considered a mere thermal HII region, Mart\'i et al. 2006) is far from
the location of the TeV source. A yet unknown pulsar wind nebula (PWN)
was proposed by Bednarek (2003), although no clear PWN signal was
observed. A distant microquasar was proposed by Paredes et al. (2007),
perhaps related with one of the X-ray/radio sources they
discovered. If such an association is accepted, the extension of the
source could be explained by the diffusion of accelerated particles
into a hypothetical nearby molecular enhancement (see Bosch-Ramon et
al. 2005).  Torres et al. (2004) and Domingo-Santamar\'ia \& Torres
(2006) studied the relationship between the TeV emission and the known
massive stars in the area, through the interaction of relativistic
protons with wind ions. The distribution of stars in the neighborhood
favors this interpretation (Butt et al. 2006). An explanation
involving the excitation of giant dipole resonances of relativistic
heavy nuclei in radiation dominated environments has also been
suggested (Anchordoqui et al. 2007).

\section{Previous very high energy $\gamma$-ray observations }

We start by making a brief summary of what has been claimed by other
experiments observing at the highest energies. 


The HEGRA IACT using four years of data (from 1999 to 2002) found a
source to the north of Cygnus X-3, steady in flux over the years,
extended, with radius 6.2$\pm$1.2$_{\rm stat}\pm$0.9$_{\rm
sys}$~arcmin, and exhibiting a hard energy spectrum with a photon
index of $\Gamma$=-1.9$\pm$ 0.1$_{\rm stat}\pm$0.3$_{\rm sys}$
(Aharonian et al. 2005). Its integral flux above 1~TeV amounts to
$\sim$5\% of the Crab Nebula, assuming a Gaussian profile for the
intrinsic source morphology. The center of the source position was
determined quite accurately at $\alpha_{\rm J2000}$=20$^{\rm h}$
31$^{\rm m}$ 57$\fs0$ $\pm$ 6$\fs2_{\rm stat}$ $\pm$ 13$\fs7_{\rm
sys}$ and $\delta_{\rm J2000}$=41$\degr$29$\arcmin$56$\farcs$8$ \pm$
1$\farcm1_{\rm stat}$ $\pm$ 1$\farcm0_{\rm sys}$.

The Whipple collaboration reported an excess at the position of the
HEGRA unidentified source (3.3$\sigma$) in their archival data of 1989
and 1990 (Lang et al. 2004), with a flux level of $\sim$ 12\% of the
Crab Nebula for E$>$600 GeV. The detected flux is in conflict with the HEGRA
flux level and steady nature of the source, assuming they all have the
same origin. This large difference between the detected flux levels,
if physical, might suggest episodic emission (with low duty cycle) or
variability over timescale measured in years. Nevertheless, the
existence of $\gamma$-ray variability is difficult to reconcile with
the extended appearance of the source. Also the large difference might
be in part due to unspecified systematic errors on the flux
determination.
Recently, the Whipple collaboration reported new observations of this
field done with their 10-m telescope for 65.5 hours during 2003 and
2005 (Konopelko et al. 2007). Their data is consistent with either a
point-like or an extended source with less than 6$^\prime$ angular
size. Regarding the position, the HEGRA and the latest Whipple data
are barely in agreement: Their centers of gravity are $\sim$9$^\prime$
apart, and only agree when adding up the spatial uncertainties in both
data sets in opposite directions. Konopelko et al. do not provide a
energy spectrum for this source, but give a 8\% Crab-level flux (although
with no energy threshold specified) under the assumption of a steep
(Crab-like) energy spectrum.

The Cygnus region shows an excess in the Milagro data (Abdo et
al. 2006). The flux at 20~TeV in a 3x3 square degree region centered
at the HEGRA position is (9.8$\pm$2.9$_{\rm stat}\pm$2.7$_{\rm sys}$)
$\times$10$^{\rm -15}$ TeV$^{-1}$ cm$^{-2}$ s$^{-1}$ assuming a
differential energy spectrum E$^{\rm -2.6}$. This flux is three times
the HEGRA flux extrapolated at 20~TeV. The Tibet Air Shower detector
recently reported evidence for an excess also in their VHE
$\gamma$-ray candidate set from this region (Amenomori et al. 2006).

In this rich observational and theoretical context we report here on
MAGIC telescope observations of TeV J2032+4130.

\section{MAGIC observations and results}

The MAGIC single dish Imaging Air Cherenkov Telescope (see e.g.,
Cortina et al. 2005 for a detailed description) is located on the
Canary Island of La Palma.
Its angular (energy) resolution is approximately 0.09$^{\rm \circ}$
(20\%), and the trigger (analysis) threshold is 55 (60) GeV at zenith
in dark conditions (see Albert et al. 2007a). One of the unique
characteristics of MAGIC is its capability of observing under moderate
Moon light illumination (Albert et al. 2007b) albeit with a slightly
elevated threshold.

The field of view of TeV~J2032+4130 was observed with MAGIC for more
than 100 hours distributed in 2005, 2006 and 2007, see
table~\ref{tab1}. During the first period in Summer 2005, the
observation was carried out in ON/OFF mode, that is, the source was
observed on-axis while observations from an empty, nearby field of
view were used to estimate the background. In Summer 2006 and 2007,
the data were taken in Wobble mode, using five positions around the
HEGRA position instead of the usual two symmetrical position in order
to monitor a wider field of view. Quality cuts based on the trigger
and after-cleaning rates were applied in order to remove bad weather
runs and data spoiled by car or satellite light flashes. After these
quality cuts the total observation time is 93.7~h. The energy range
for which we report these results is significantly above the
aforementioned trigger and analysis threshold energies due to the fact
that the observations were scheduled during moonlight and at
relatively high zenith angles (up to 44$^o$).

The data analysis was carried out using the standard MAGIC analysis
and reconstruction software (Bretz \& Wagner 2003). It follows the
general stream explained in Albert et al.  (2006b,c,d). After
calibration and two levels of image cleaning tail cuts (for image core
and boundary pixels, see Fegan 1997), the camera images are
parameterized by the so-called image parameters (Hillas 1985). The
Random Forest method was applied for the $\gamma$/hadron separation
(Albert et al. 2007c). Using this method a parameter, dubbed
hadronness (H), can be calculated for every event and which is a
measure of the probability that the event is not $\gamma$ like. The
$\gamma$ like sample is selected for images with a H below a specified
value, which is optimized using a sample of Crab Nebula data processed
with the same analysis stream. An independent sample of Monte Carlo
$\gamma$-showers was used to determine the cut efficiency.
Since part of our observations was recorded during partial moon-shine,
we have corrected the efficiency loss due to the increase of ambient
light following the procedure outlined in Albert et al. (2007b).

\begin{figure}[!t]
\centering
\includegraphics*[angle=0,width=0.9\columnwidth]{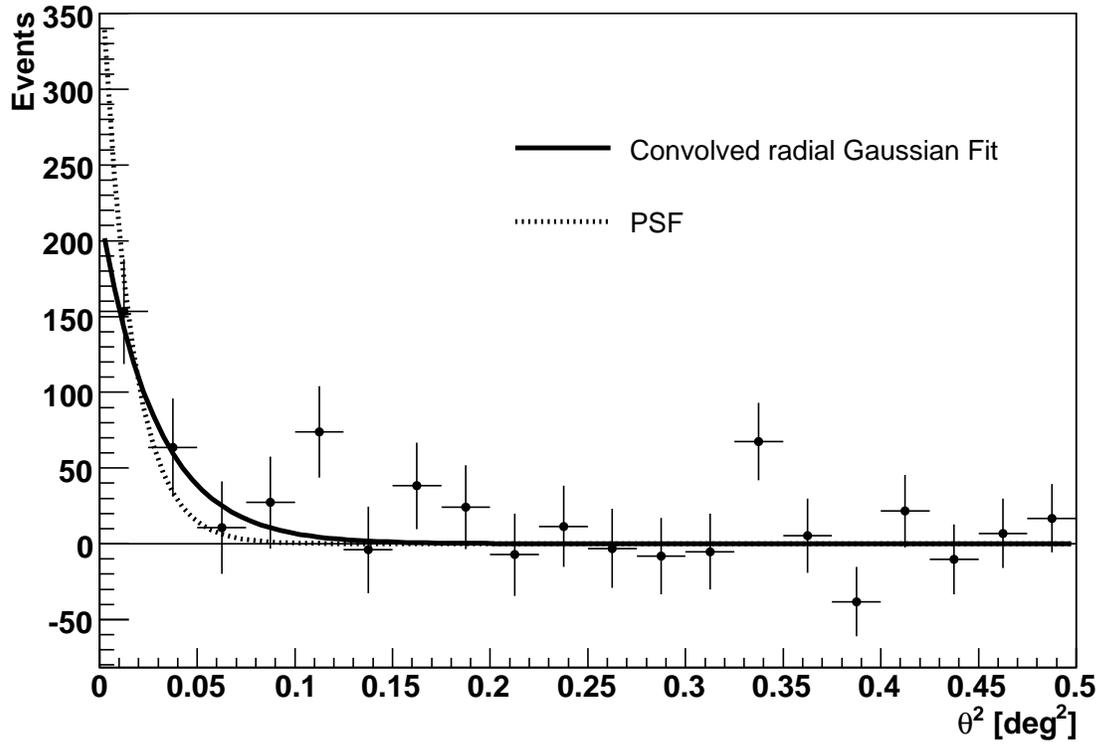}
\caption{Distribution of the $\theta^2$-parameter for events coming from the direction of TeV J2032+4130 (SIZE$>$800~pe), the background distribution subtracted (black points). A convolved radial Gaussian fit F=A$\times$exp(-0.5$\theta^2$/($\sigma_{psf}^2 +\sigma_{src}^2$)) is indicated by the solid black line with $\sigma_{src}=5.0$$\pm$1.7 arcmin. The $\sigma_{psf}$ was measured from MC simulation and validated with Crab Nebula observations to be $\sigma_{psf}$=5.2$\pm$0.1 arcmin (dashed black line).}
\label{fig1} 
\end{figure}

The $\theta^2$-distribution was calculated, being $\theta$ the angular
distance between the source direction and the reconstructed arrival
direction of the showers. The reconstruction of individual
$\gamma$-ray arrival directions makes use of the DISP method
(Domingo-Santamaria et al. 2005). The expected number of background
events is calculated using five regions symmetrically placed for each
wobble position with respect to the center of the camera and refered
to as anti-sources. Figure \ref{fig1} shows the distribution of the
$\theta^2$ parameter for the excess observed from the direction of the
source, for a SIZE cut of 800 photoelectrons (pe). This relatively
high SIZE cut was selected in order to optimize the sensitivity for a
source with such a hard energy spectrum observed during
moonlight. Therefore, the total number of $\gamma$-like excesses after
Hillas cuts and applying a cut in $\theta^2<$0.05, is $N_{\rm
ex}=233$, for which a total significance of $5.6\sigma$ is
obtained. Table~\ref{tab2} shows the number of excesses above
background for the different observing periods.

The excess is fitted to a Gaussian function folded with the telescope
PSF, as obtained from Monte Carlo simulations and validated with Crab
Nebula observations. The source is extended with respect to the MAGIC
PSF. Its intrinsic size assuming a Gaussian profile is $\sigma_{\rm
src}$ =5.0$\pm$1.7$_{\rm sta}\pm$0.6$_{\rm sys}$ arcmin. The exact
shape of the source, even if similar to the keV diffuse emission
reported by Horns et al. 2007, cannot be completely trusted due to
limited statistics and telescope pointing systematics.

\begin{figure}[!t]
\centering
\includegraphics*[angle=0,width=0.9\columnwidth]{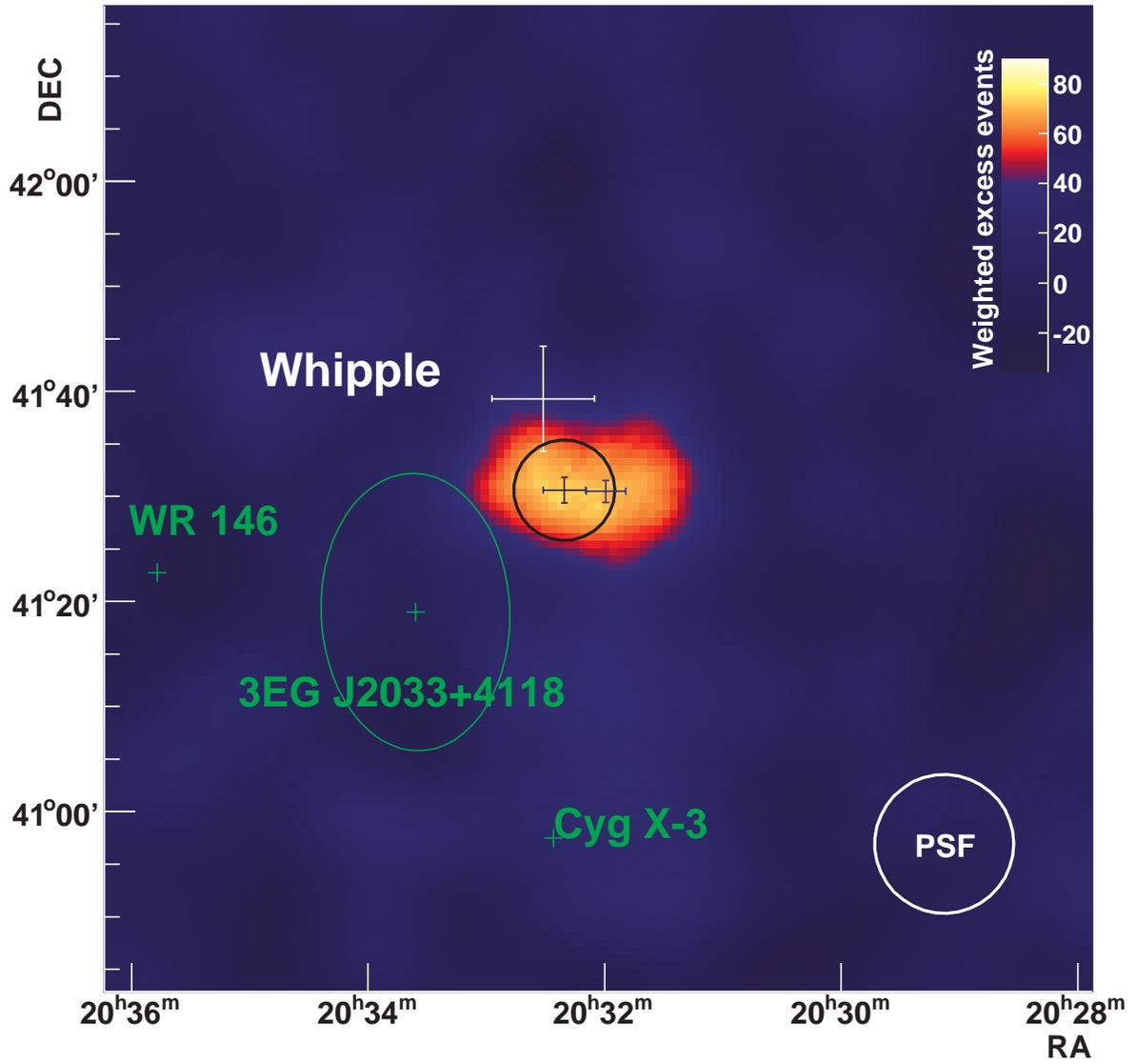}
\caption{Gaussian-smoothed ($\sigma$=4') map of $\gamma$-ray excess events
(background-subtracted) for energies above 500~GeV. The MAGIC position
is shown with a black cross. The surrounding black circle corresponds
to the measured 1$\sigma$ width. The last position reported by Whipple
is marked with a white cross while the HEGRA position is shown with a
blue cross in the center of the field of view. The error bars, in all
cases, correspond to the linear sum of the statistical and systematic
errors. The green crosses correspond to the positions of Cyg X-3,
WR~146 and the EGRET source 3EG~J2033+4118. The ellipse around the
EGRET source marks the 95$\%$ confidence contour.}
\label{fig2} 
\end{figure}

Figure \ref{fig2} shows the Gaussian-smoothed ($\sigma$=4') map
(0.65$^{o}\times$0.65$^{o}$) of $\gamma$-ray (background subtracted)
around TeV J2032+4130 for energies E $>$ 500~GeV. The position of a
few previously observed $\gamma$-ray source candidates are also shown,
namely Cyg X-3, the EGRET source 3EG~J2033+4118 (with its confidence
contour at 95$\%$), the Wolf Rayet star WR~146, and the Whipple and
HEGRA experimental positions. The regions around Cyg X-3, WR~146 and
3EG~J2033+4118 have been further investigated by us and no detection
is obtained for a steady emission. The upper limit fluxes (Rolke et
al. 2005) for 95$\%$ confidence level, above 500 GeV for a point-like
source at these positions are given in table \ref{tab3}.

To determine the best position of the MAGIC detection the excess map
was fitted to a 2D bell-shaped function. The result is shown in the
skymap with a black cross as well as by a circle indicating its
size. The best-fit coordinates are RA$_{\rm J2000}$=20$^{\rm h}$
32$^{\rm m}$ 20$^{\rm s}$ $\pm$ 11$^{\rm s}_{\rm stat}$ $\pm $11$^{\rm
s}_{\rm sys}$ and DEC$_{\rm J2000}$=41$\degr$ 30$\arcmin$ 36$\farcs$0
$\pm$ 1$\farcm2_{\rm stat}$ $\pm$ 1$\farcm8_{\rm sys}$ (for more
details on the systematic uncertainties in the source position
determination, see Bretz et al. 2003). The position found is
compatible within errors with the one determined by HEGRA, and barely
compatible with the claims by Whipple mentioned above (in Konopelko et
al. 2007).

\begin{figure}[!t]
\centering
\includegraphics*[angle=0,width=\columnwidth]{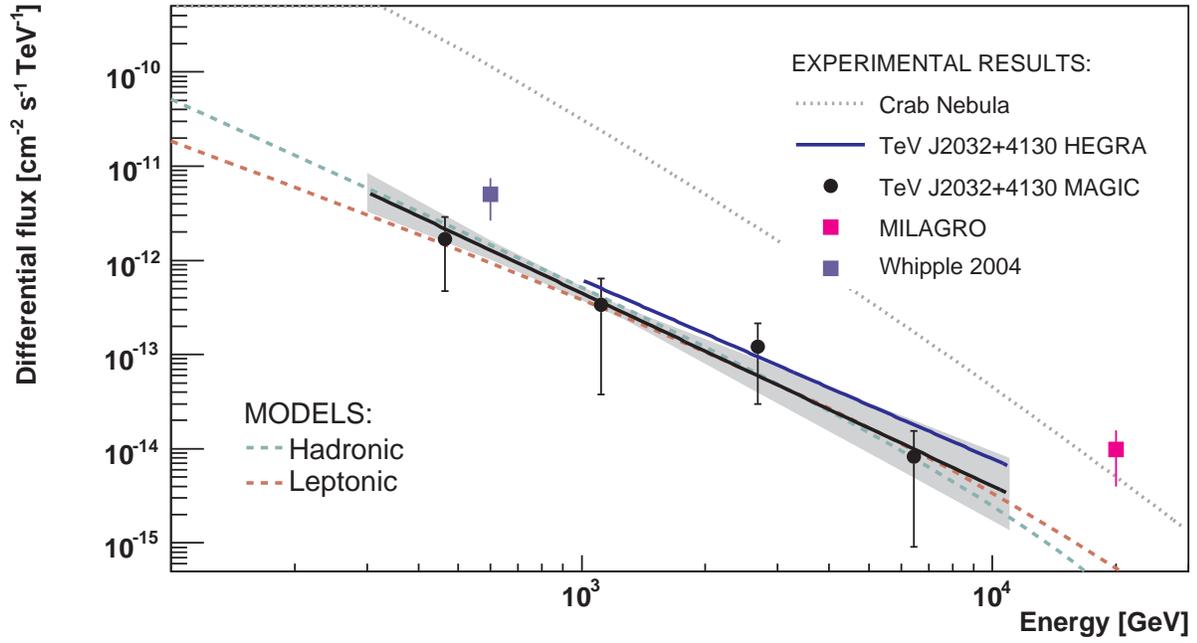}
\caption{Differential energy spectrum from TeV J2032+4130 as measured by the MAGIC telescope in black solid line. The grey shadow shows the 1$\sigma$ error in the fitted energy spectrum. The flux observed by Whipple in 2005 and in the Milagro scan are marked with colored squares (blue and pink, respectively). The grey dotted line represents the Crab Nebula energy spectrum measured by MAGIC. The blue line shows the HEGRA energy spectrum. Theoretical one-zone model predictions are depicted with dashed lines.}
\label{fig3} 
\end{figure}

The TeV J2032+4130 energy spectrum was obtained using the Tikhonov
unfolding technique (Tikhonov $\&$ Arsenin 1979). It can be fitted
($\chi^2/n.d.f=0.3$) by a power law function. The differential flux
(TeV$^{-1}$cm$^{-2}$s$^{-1}$) is:
\begin{equation}
\frac{dN}{dE dA dt} = (4.5\pm 0.3)\times10^{-13}(E/1~TeV)^{-2.0\pm0.3} 
\end{equation}
The errors quoted are only statistical. The systematic error is
estimated to be 35$\%$ in the flux level and 0.2 in the photon index
(see Albert et al. 2007a). The differential energy spectrum is shown
in Figure \ref{fig3}. The HEGRA TeV J2032+4130 and MAGIC Crab Nebula
measured spectra (in Albert et al. 2007a) are shown in blue solid line
and black dotted line, respectively. The MAGIC energy spectrum is
compatible both in flux level and photon index with the one measured
by HEGRA.

Crab Nebula data from the same periods and zenith angle distributions
were studied with the same analysis chain to check for any systematic
deviation due to the long observation period. No indication of time
variability was observed: the source integral flux is constant within
errors, at 3$\%$ of the Crab Nebula flux. The relative systematic
uncertainty in the ratio of both fluxes was estimated to be less than
10$\%$. This uncertainty comes mainly from the slightly different
atmospheric transmission conditions and differences in the detector
parameters during data taking of the source and the Crab Nebula.

For illustrative purposes, the dotted lines in Figure \ref{fig3}
represent one-zone hadronic and leptonic models of the high energy
emission, both consistent with observations at lower energies in the
region. Under the hadronic scenario, the $\pi^0$ are obtained from a
proton parent population described by a power law (index $\Gamma=-2$)
with exponential cutoff at 100 TeV. The cutoff value was adopted to be
consistent with the upper limit at the highest energies coming from
the HEGRA spectrum. The inverse Compton spectrum is obtained from an
electron population with equal index and a 40 TeV exponential cutoff
scattering off the CMB photons. As in Aharonian et al. 2005, we do not
consider here the conditions under which particles are accelerated or
how they lose energy. Our leptonic fits (see also the quoted paper for
an SED representation) can only cope with the data if we are actually
looking at a Compton peak around the energy range of detection, which
is not fully discarded within errors. Both models are compatible with
the high energy emission. Confirming the reality of the diffuse
emission detected at lower energies is crucial to distinguish between
these and more complex models.


\section{Concluding remarks}

MAGIC observations confirm the location of TeV J2032+4130 found by
HEGRA. The MAGIC observation shows an extended source with a
significance of 5.6$\sigma$. We find a steady flux with no significant
variability within the three year span of the observations (with the
flux being at a similar level of the HEGRA data of the period
2002-2005). We also present the source energy spectrum obtained with
the lowest energy threshold to date, which, within errors, is
compatible with a single power law.


\acknowledgements

We thank the IAC for the excellent working conditions at the ORM. The
support of the German BMBF and MPG, the Italian INFN, the Spanish
CICYT, the ETH Research Grant TH 34/04 3, and the Polish MNiI Grant
1P03D01028 is gratefully acknowledged.

\clearpage

\begin{table}
\caption{Observing periods, zenith angle ranges and observation modes.}
\begin{center}
\vspace{-.45cm}
\begin{tabular}{llll}
\hline
\hline	
Year & T[h] & Z.A.[deg]  & Mode \\
\hline
2005 & 18.1 & 13--30 & ON/OFF\\
2006 & 60.1 & 11--44 & Wobble\\ 
2007 & 15.5 & 11--30 & Wobble\\ 
\hline
\end{tabular}
\label{tab1}
\end{center}
\end{table}

\clearpage

\begin{table}
\caption{Events recorded above SIZE$>$ 800~pe. $N_{\rm on}$, $N_{\rm off}$ and $N_{\rm ex}$ refer to the number of events recorded in the direction of the source, the normalized background and the $\gamma$-ray excess respectively. The normalization ratio $f_{\rm norm}$ and significance $N_{\rm \sigma}$ are also shown.}
\begin{center}
\vspace{-.45cm}
\begin{tabular}{lccccc}
\hline
Year & $N_{\rm on}$ & $N_{\rm off}$ & N$_{\rm ex}$ & $f_{\rm norm}$ &
$N_{\rm \sigma}$
\\
\hline
2005 & 641 & 576 &  65 & 0.47 & 2.2\\
2006 & 688 & 559 & 129 & 0.20 & 4.8\\
2007 & 175 & 136 &  39 & 0.20 & 2.9 \\
Overall & 1504 & 1271 & 233 &0.27 & 5.6\\
\hline
\end{tabular}	
\end{center}
\label{tab2}
\end{table}

\clearpage

\begin{table}[!t]
\caption{UL for sources in the FOV, above 500 GeV.}
\vspace{-.45cm}
\begin{center}
\begin{tabular}{ll}
\hline
\hline
Target Name & Flux (Crab Nebula) \\
\hline
Cyg X-3 & 0.011 \\ 
WR~146 & 0.010  \\ 
3EG~J2033+4118 & 0.009 \\
\hline
\end{tabular}
\label{tab3}
\end{center}
\end{table}

\clearpage

\end{document}